# PatchSorter: A High Throughput Deep Learning Digital Pathology Tool for Object Labeling

Cedric Walker[1], Tasneem Talawalla[2], Robert Toth[3], Akhil Ambekar[4,5], Kien Rea[2], Oswin Chamian[2], Fan Fan[2], Sabina Berezowska[6], Sven Rottenberg[1,7], Anant Madabhushi[8,9], Marie Maillard[6], Laura Barisoni[4,10], Hugo Mark Horlings[11], Andrew Janowczyk[8,12,13]

[1] Institute of Animal Pathology, Vetsuisse Faculty, University of Bern, Bern, Switzerland
[2] Department of Biomedical Engineering, Case Western Reserve University, OH, USA
[3] Toth Technology LLC, Toth Technology LLC, NJ, USA
[4] Department of Pathology, Division of AI & Computational Pathology, Duke University, NC, USA
[5] AI Health, Duke University, NC, USA
[6] Institute of Pathology, Lausanne University Hospital, Vaud, Switzerland
[7] Bern Center for Precision Medicine, University of Bern, Bern, Switzerland
[8] Department of Biomedical Engineering, Emory University and Georgia Institute of Technology, GA, USA
[9] Atlanta Veterans Medical Center, Atlanta, GA, USA
[10] Department of Medicine, Division of Nephrology, Duke University, Durham, NC, USA
[11] Department of Pathology, The Netherlands Cancer Institute, Amsterdam, The Netherlands
[12] Department of Oncology, Division of Precision Oncology, University Hospital of Geneva, Geneva, Switzerland
[13] Department of Clinical Pathology, Division of Clinical Pathology, University Hospital of Geneva, Geneva, Switzerland

## Abstract

The discovery of patterns associated with diagnosis, prognosis, and therapy response in digital pathology images often requires intractable labeling of large quantities of histological objects. Here we release an open-source labeling tool, PatchSorter, which integrates deep learning with an intuitive web interface. Using >100,000 objects, we demonstrate a >7x improvement in labels per second over unaided labeling, with minimal impact on labeling accuracy, thus enabling high-throughput labeling of large datasets.

## Main

The increasing digitization of routine clinical histology slides into whole slide images (WSI) has spurred great interest in the development of WSI-based biomarkers for diagnosis, prognosis, and therapy response [1–3]. These biomarkers are typically based on patterns associated with the location and type of individual histologic objects (e.g., cells – lymphocytes/epithelial; glomeruli – globally sclerotic (GS)/non-sclerotic (non-GS/SS)/segmentally sclerotic (SS); tubules – distal/proximal; tumor buds – present/absent). While current hardware and machine learning algorithms can locate and type objects at scale, the manual assignment and review of large labeled datasets used to train or validate models remains arduous. For example, a single WSI may contain over 1 million cells, which, if requiring a modest 1 second per cell to label, would result in approximately 12 non-stop days of effort. To aid experts (e.g., pathologists) in this labeling process, several image analysis algorithms have been proposed[4–9]. However, these algorithms tend to either (a) not be integrated into polished, user-friendly tools, making them unsuitable for usage by domain experts, or (b) are of a closed source, for-profit nature, creating a barrier to their broad-usage, which potentially limits their continuous improvement via the facile integration and evaluation of new algorithms[10].

Appreciating the need for an open-source force multiplier for labeling histological objects, we here describe and make available to the community PatchSorter (PS). PS is a user-friendly, browser-based tool, which allows the user to leverage deep learning (DL) to quickly review and apply labels at a group, as opposed to a single object, level (**Figure 1**). We demonstrated that this "bulk" labeling approach improves labeling efficiency across four use cases, spanning three levels of increasing object complexity (*i.e.*, objects comprised of increasing number of cells and cell types) (**Table 1**).

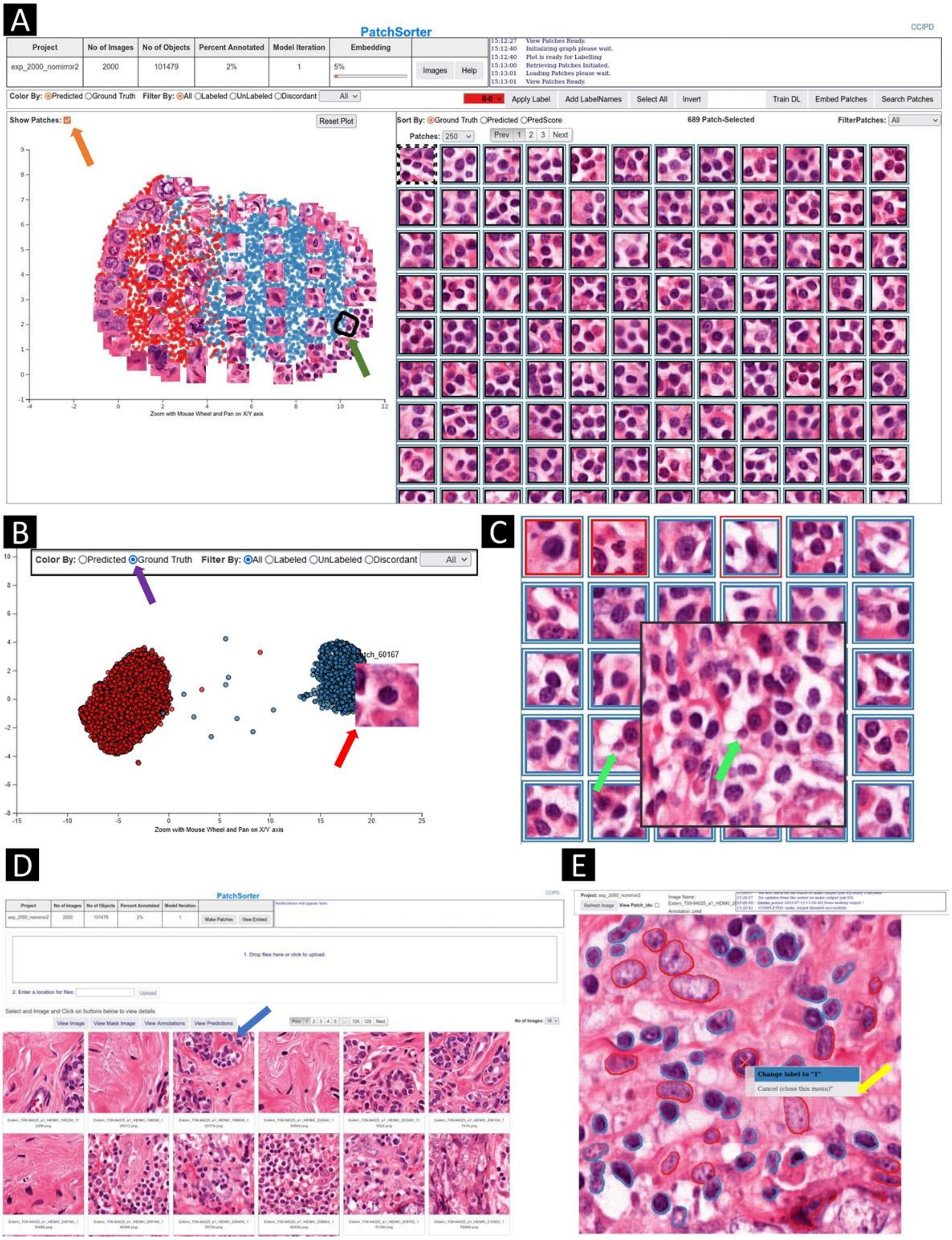

**Figure 1.** PatchSorter user interface. (A) The embedding plot after initial embedding (left) with corresponding grid plot (right). The two-dimensional embedding plot places patches with the same deep learned features in close proximity, causing objects with the same object class to cluster. The user lassos points (black contour with green arrow) which then appear in the grid plot for labeling using efficient keyboard shortcuts. In the embedding plot, a subset of patches can be overlaid to aid in selecting regions in the embedding space (orange arrow). (B) The embedding plot allows for coloring patches by prediction and ground truth (purple arrow). The embedding plot shows the same data set as (A) after eight model iterations where the embedding space is well separated by ground truth labels. Hovering over a point in the embedding space shows the corresponding patch (red arrow). (C) Grid plot coloring shows current predictions and ground truth. The inner square color represents ground truth while the outer square color represents model prediction, with black indicating that the patch is not yet labeled. Right-clicking on patch in the grid plot shows a larger region of interest (ROI) for context (green arrows). (D) From the image pane, prediction and ground truth labels can be visualized (blue arrow) in the output reviewer. (E) Here, objects labels can be updated via a right click on the object (yellow arrow).

PS enables labeling speed improvements by using DL derived features to embed patches containing the object of interest (e.g., glomeruli) into a 2-dimensional embedding space, such that similarly presenting objects are proximally located**.** The user then reviews patches within a localized region that are likely to correspond to the same class, thus enabling assignment of labels in bulk (*i.e.*, assignment of the same label to multiple objects at once) with increased efficiency. The DL model and associated embedding space is then iteratively refined with the user's feedback, yielding improved class separability, further improving subsequent labeling efficiency (**Figure S1** and **S2**).

To evaluate this improved efficiency, a labels per second (LPS) metric was compared between PS and an unaided approach, Quick Reviewer (QR, see **Online Methods**)[11], across four use cases (**Table 1**, see **Online Methods**) totaling over 120,000 objects. QR was used to label a random subset of the data to estimate manual LPS ($M_t$) per use case. Efficiency improvement was measured as the ratio ($\theta_t$) between PS's LPS ($PS_t$) and $M_t$. To ensure labeling efficiency improvements did not come at the cost of label fidelity, concordance between QR and PS assigned labels was measured. Labeling for all use cases was conducted by board-certified pathologists, after having received an introduction to the PS and QR user interfaces.

| Level of cellular and structural complexity | Histological Primitive | Number of ROI | Number of histologic objects | PS manual time (s) | PS human time (s) | PS total time (LPS) | PS human time ($PS_t$, LPS) | Manual time ($M_t$, LPS) | Speed up ($\theta_t$) | PS interval low (LPS) with speed up | PS interval high (LPS) with speed up | Stain type | Concordance |
|---|---|---|---|---|---|---|---|---|---|---|---|---|---|
| Low | Breast cancer nuclei: **lymphocytes** vs. non-lymphocytes | 2000 | 101479 | 235998 | 32735 | 1.92 | 3.1 | 0.43 | 7.21x | 0.35 (0.81x) | 9.6 (22.3x) | H&E | 96% |
| Medium | Lung cancer **tumor-budding**: present vs. absent | 27 | 1631 | 2471 | 1800 | 0.66 | 0.906 | 0.292 | 3.1x | 0.49 (1.68x) | 1.06 (3.6x) | H&E | 93% |
| Medium | Kidney **tubules**: distal vs. proximal vs. abnormal | 216 | 2298 | 10943 | 3648 | 0.51 | 0.63 | 0.21 | 2.89x | 0.52 (2.47x) | 0.95 (4.5x) | PAS | 97% |
| High | Kidney **glomeruli**: SS vs. GS vs non-SS/GS | 16158 | 16158 | 23978 | 20171 | 0.673 | 0.801 | 0.159 | 5.03x | 0.57 (3.58x) | 1.15 (7.23x) | PAS | 96% |

**Table 1.** Description of the datasets used for validating PatchSorter along with the demonstrated efficiency gains in terms of labels per second and concordance with an unaided approach. The difference between human time and total time is the inclusion of model training and embedding in the lableing time in total time, while it is removed for human time, as the human reader can be dismissed to perform other non-labeling related tasks. Manual time for the same task is estimated based on the extrapolatation of manual labeling of a subset of the data. Upper and lower intervals for PatchSorter human time are estimated by measuring $PS_t$ in 15-minute intervals, corresponding to total QuickReviewer labeling time. For the nuclei use case, speed increases of up to 22.3x (9.6 LPS) are observable while only being slightly slower than manual labeling in one of the measured 15 minute intervals. For tubules, tumor buds and glomeruli, PatchSorter offers a speed increase over manual labeling efforts, even for worst-case estimates. SS = segmentally sclerotic, GS = globally sclerotic, non-SS/GS = non-sclerotic.

These results indicate that (a) PS provides sizable efficiency improvements in labeling objects of all levels of cellular and structural complexity, while (b) not coming at the cost of a loss of labeling accuracy (**Table 1**). Interestingly, differences remain in labels generated via PS and QR. This difference can be at least partially attributed to label uncertainty related to ambiguous objects, wherein labeling is likely to suffer from inter/intra observer variability (**Figure S3-S6**).

The usage of PS appears to proceed in two distinct workflows: (a) rapid bulk labeling on the periphery of the embedding space where objects with more obvious labels tend to be grouped, and (b) slower intricate labeling at the interface between classes where object labels tend to be more challenging to determine. Notably, these challenging data points often drive improved class separation. As such, our suggested best practice is to alternate between the two workflows: (1) when class separation is high in the embedding plot the operator should focus on bulk labeling, while (2) if class separation is low, labeling should be performed at the interface between classes. This interface labeling should result in improved class separation in the next embedding iteration, thus facilitating again bulk labeling (**Figure S1**).

The transition point between these two workflows appears to be use case specific (**Figure 2**). While in the nuclei use case labeling speed improves with DL training, in the glomerular use case, a more time-consuming careful evaluation is required throughout the task, due to the difficult nature of differentiating between transitioning classes (e.g., SS with small areas of scarring mimicking non-GS/SS or with extensive segmental sclerosis mimicking GS).

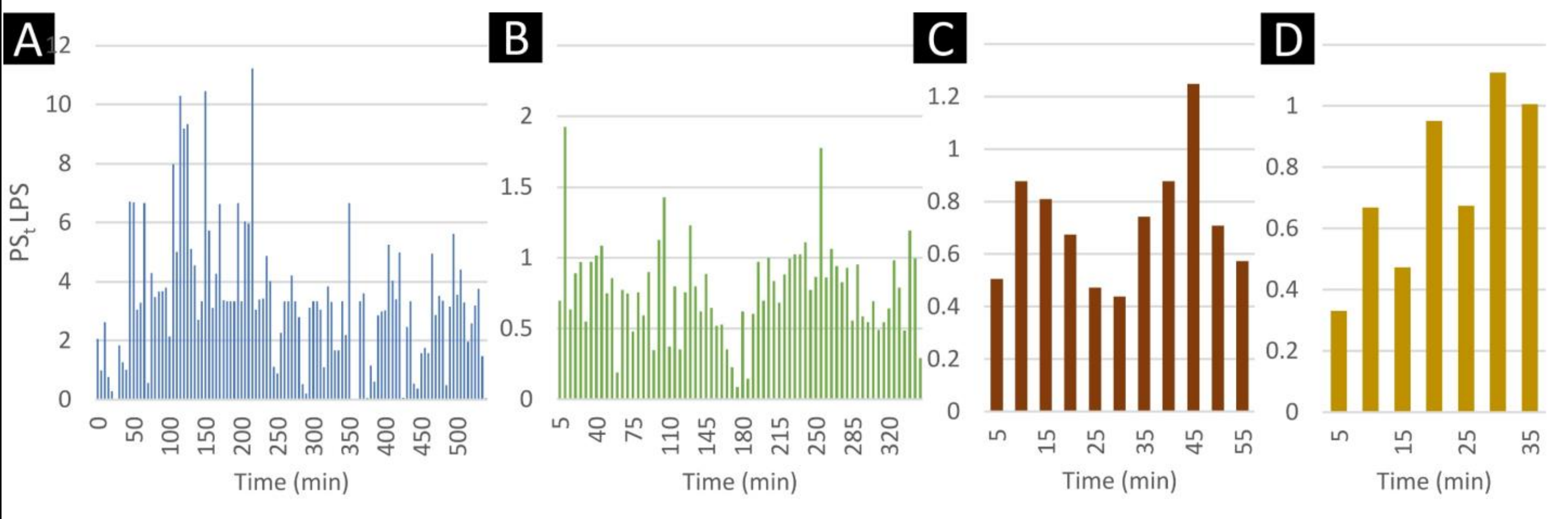

**Figure 2.** Efficiency metric $PS_t$ over time measured in 5-minute intervals demonstrating the improvement in labeling speed of PS for (A) nuclei, (B) glomeruli, (C) tubules, and (D) tumor bud use case. The x-axis is the human annotation time in minutes and the y-axis is the labeling speed per second for a given time interval. Performance improvement over time varies per use case. For (A) nuclei labeling, a consistent performance increase is noted, consistent with increased class separation. As the entire dataset is labeled, performance decreases as easy-to-discern object labels are exhausted. For (B) glomeruli labeling, the initial embedding allowed for bulk annotation of non-SS/GS, GS and SS at the edge of the embedding plot, while later, nuanced labeling had to be employed due to the task's difficulty. For (C) tubule labeling, the initial embedding allows for bulk annotation. While due to changes to the initially assigned labels and imbalanced labeling of the 4 classes, class separation decreased in the subsequent iteration, further iterations led to increased class separability and labeling efficiency. Lastly, for (D) tumor bud candidates, initial labeling efficiency was only marginally higher than manual baseline LPS. As class separability increased, so did labeling efficiency.

From a usage perspective, after PS installation, no internet connection is required, enabling its use in clinical environments where data may not be anonymized. PS can be installed locally on commodity desktops or deployed on servers for remote access by experts (*i.e.*, bringing the expert to the data), as datasets become too large to quickly transfer and clinical environments further restrict the installation of 3rd party software. While PS has been validated in this study on hematoxylin and eosin (H&E) and periodic acid-Schiff (PAS) staining, given the DL-based back end, PS can be considered agnostic to stain type and be used with any stain, image, or object type.

In conclusion, PS is a user-friendly, high-throughput object labeling tool being publicly released for community usage, review, and feedback. PS has demonstrated significant improvement in efficiency in object labeling in the hands of domain experts without sacrificing labeling accuracy. The source code of PS is freely available for use, modification, and contribution at www.patchsorter.com.

## Citations

**Acknowledgments**

Research reported in this study was supported by the NIH under awards U01CA239055, 1R01LM013864, 1U01DK133090, U01CA248226, along with NIDDK/NIH under the award number 2R01DK118431-04 and the Nephcure kidney international foundation. The Nephrotic Syndrome Study Network (NEPTUNE) is part of the Rare Diseases Clinical Research Network (RDCRN), which is funded by the National Institutes of Health (NIH) and led by the National Center for Advancing Translational Sciences (NCATS) through its Division of Rare Diseases Research Innovation (DRDRI). NEPTUNE is funded under grant number U54DK083912 as a collaboration between NCATS and the National Institute of Diabetes and Digestive and Kidney Diseases (NIDDK). Additional funding and/or programmatic support is provided by the University of Michigan, NephCure Kidney International and the Halpin Foundation. RDCRN consortia are supported by the RDCRN Data Management and Coordinating Center (DMCC), funded by NCATS and the National Institute of Neurological Disorders and Stroke (NINDS) under U2CTR002818.

We are also grateful to the Vetsuisse Faculty (University of Bern) for their support of digital pathology. Moreover, financial support came from the European Union (ERC AdG-88387) and the Department of Defense Ovarian Cancer Research Program [W81XWH-22-1-0557].

Funding for the CureGN consortium is provided by U24DK100845 (formerly UM1DK100845), U01DK100846 (formerly UM1DK100846), U01DK100876 (formerly UM1DK100876), U01DK100866 (formerly UM1DK100866), and U01DK100867 (formerly UM1DK100867) from the National Institute of Diabetes and Digestive and Kidney Diseases (NIDDK). Patient recruitment is supported by NephCure Kidney International.

The MATADOR study is supported by the Dutch Cancer Society (CKTO 2004-04) and unrestricted research grants from Sanofi and Amgen. The funders had no role in the study design; in the collection, analysis, and interpretation of data; in the writing of the paper; or in the decision to submit the paper for publication. We would like to thank all patients and their families, to the MATADOR study teams of the participating centers, the Data Center of the Netherlands Cancer Institute for collecting the data, the Core Facility – Molecular Pathology and Biobanking of the Netherlands Cancer Institute, the Dutch Breast Cancer Research Group (BOOG) for their role in coordinating the study, the Dutch Cancer Society and of the Dutch Ministry of Health, Welfare and Sport for their funding and Sanofi and Amgen for their unrestricted research grants.

**Conflict of Interest Statement**

AM is an equity holder in Picture Health, Elucid Bioimaging, and Inspirata Inc. Currently he serves on the advisory board of Picture Health, Aiforia Inc, and SimBioSys. He also currently consults for SimBioSys. He also has sponsored research agreements with AstraZeneca, Boehringer-Ingelheim, Eli-Lilly and Bristol Myers-Squibb. His technology has been licensed to Picture Health and Elucid Bioimaging. He is also involved in 3 different R01 grants with Inspirata Inc. LB is a consultant for Sangamo and Protalix and is on the scientific advisory boards of Vertex and Nephcure. AJ provides consulting for Merck, Lunaphore, and Roche, the latter of which he also has a sponsored research agreement. HMH received financial compensation from Roche Diagnostics BV paid to the institute. No other conflicts of interest were declared.

**Contributing Author Statement**

C.W. contributed to the study conceptualization and design, data curation, interpretation of data, experiment execution and drafting of the manuscript; A.A., K.R., O.C, F.F., S.B., M.M. L.B. and H.M.H. contributed to the study design, data curation, interpretation of data, and experiment execution; T.T. and R.T. contributed to the code base, study conceptualization and study design; A.J. contributed to the methodology, code base, study conceptualization and design, gathering of resources, data curation, interpretation of data, reviewing, and editing; S.R. and A.M. contributed to study conceptualization, interpretation of data, reviewing, and editing.

## Online Methods

### PatchSorter workflow

As per PS workflow (**Figure S2),** images were uploaded to PS together with a corresponding segmentation mask highlighting object location. PS then extracts patches, with user configurable patch sizes, around the center of these objects to create an internal database for high-speed training. While a number of different self-supervised approaches are supported by PS (e.g., BarlowTwins[12], and AutoEncoder[13]), a SimCLR[14] using a ResNet18[15] backbone was trained using contrastive loss, creating a dataset specific DL feature space. Feature vectors are computed for each patch using this learned feature space, and are subsequently embedded using Uniform Manifold Approximation and Projection (UMAP)[16] into 2-dimensions. As a result of this process, objects which look the same tend to be plotted near each other in the embedding plot. This allows the user to lasso regions on the embedding plot and provide the label for the selection in the grid plot (**Figure 1A**). As more objects are labeled, PS is increasingly able to learn a more discriminative feature space for the categories of the specific task. As a result, subsequent iterations should demonstrate improved localized clustering "purity" (*i.e.*, objects in the same cluster have the same label). This approach has two consequences, (a) the user can avoid intractably manipulating individual objects and instead provide bulk annotations to groups of objects with a single input, and (b) as the DL model (and thus the embedding space) is refined with the user's feedback, the user can begin to see regions in the 2d space, where the underlying model is struggling to differentiate between class-types. The visibility of such regions affords the user the opportunity to better invest their time in selecting objects that when labeled are most likely to further improve class separability in the next iteration, which in turn further improves subsequent labeling efficiency.

To facilitate the efficiency of this bulk labeling process, features from modern operating systems were implemented, such as drag-select and numerous intuitive keyboard shortcuts for (a) selecting all objects, (b) inverting the selection, as well as (c) changing the desired label (e.g., "1" selects the first class). If specific objects of interest are sought, PS provides content-based image retrieval, wherein the user may upload a patch of the object of interest, and similar objects from the dataset will appear for labeling within the standard workflow. PS was designed in a decoupled, modular, manner such that its backend technologies can easily be exchanged to evaluate different DL technologies, with minimal modifications to the base application. To ease integration with other workflows and pipelines, the output of PS is highly portable: mask images with color indicating class membership (**Figure S1D**). For more advanced users, the internal database can be directly employed in common downstream tasks, such as training large custom DL models. It is important to note, that the user retains full control over the accuracy of object labels at all times, and only confirmed labels are stored. Usefully, these newly generated ground truth labels (as well as predicted labels), can be visualized through PS for rapid tile-level review, wherein individual object labels may still be modified as needed (**Figure 1E**).

### Manual unaided baseline efficiency estimation

Quick Reviewer (QR)[11], an open-source object labeling tool, was employed as the unaided baseline approach for comparison against PS. QR is a simple web-based framework which presents an image patch to the user, one at a time, and collects their label determination via a button click. It should be noted that QR already offers notable efficiency advantages over true unaided manual object labeling pipelines, as objects are directly presented to the user, which obviates the time-consuming process of (a) finding specific objects in WSIs, and (b) transitioning between different WSIs. As such, QR times can be considered optimistic as compared to a "fully" unaided approach, which are increasingly becoming less common in practice.

### Metrics for evaluating PS efficiency improvement

For comparing PS to QR we introduce a labels per second (LPS) metric. For each of the 4 use cases described below, QR was used to label a random subset of the data to estimate LPS and extrapolate manual LPS ($M_t$) for the entire dataset. For PS, we measure LPS in total time and human time ($PS_t$). The difference between human time and total time is the inclusion of model training and patch embedding in total time, while it is removed for human time, as the the human reader can be dismissed to perform other non-labeleing related tasks. Efficiency improvement is then measured as the ratio ($\theta_t$) between $PS_t$ and $M_t$. To ensure these labeling efficiency improvements did not come at the cost of unacceptable fidelity loss, the subset of data manually labeled is quantitively compared using the concordance metric to the labels produced via PS.

**Use Case 1: Lymphocyte labeling in triple-negative breast cancer**

Tumor infiltrating lymphocytes (TILs) have emerged as a biomarker of interest in breast cancer, with mounting evidence of their prognostic and predictive value in triple-negative breast cancer[17]. TILs are labeled in accordance with the immune oncology working group guidelines for immune infiltration scoring in breast cancer[18] into lymphocyte and non-lymphocyte.

To begin, 2000 1000x1000 pixel image tiles were randomly cropped from n=21 fully deidentified H&E WSIs scanned at 40x Magnification from the MATADOR[19] cohort, ensuring sufficient quality (e.g., exclusion of tissue folds or blurry regions). ROIs were stain normalized based on a reference tile from the MATADOR[19] cohort using the Vahadane stain normalization[20] implementation from StainTools (https://github.com/Peter554/StainTools). Using the HoverNet[21] implementation from histocartography[22], nuclei were segmented to provide the object location information to PS. Following the PS workflow (**Figure S2**), ROIs and corresponding object segmentation mask were uploaded into PS where nuclei were extracted from the ROI into 64x64 pixel patches with the nuclei centered.

**Use Case 2: Detection of tumor budding in pulmonary squamous cell carcinoma**

Tumor buds, defined as clusters of cancer cells composed of fewer than five cells[23], is an invasive pattern that has been described in solid tumors (e.g., colon cancer). Tumor budding has attracted interest as a prognostic biomarker in lung cancer, with the presence of tumor buds being associated with worse patient outcome.

Here, 27 2000x2000 pixel ROI were extracted at 40x from n=3 fully deidentified H&E stained lung cancer samples. A u-net[24] model was applied to each ROI to segment potential tumor bud candidates for further labeling into absent/present. ROI were stain normalized using the Vahadane stain normalization[20] method implemented in StainTools and each ROI was downsampled to 500 by 500 pixel using nearest-neighbor interpolation. 1761 tumor bud candidates were extracted into 64x64 pixel patches by PS with a single potential tumor bud centered.

Small changes to the PS user interface were made to show a larger 256x256 image instead of the 64x64 image used for training the DL model. This provided additional context was requested by the reader to improve their decision-making comfort; these changes are available in the PS code repository. In QR, patches were presented with an overlay of the u-net segmentation mask for indicating tumor bud position in the ROI, as multiple tumor buds might be present in the ROI.

In addition to absent/present, PS and QR were set-up to include an 'unsure' label, allowing for the labeling of patches where the pathologist was not comfortable in making a definitive decision during the experiment. The reported accuracy is measured between all labels present in QR and PS (absent/present/unsure).

Discussion of the discordant cases between QR and PS indicated that the additional context provided by QR led the pathologist to be less confident in labeling patches as 'absent', while in PS, patch similarities to other 'absent' examples in the embedding space led the pathologist to more likely label these patches as 'absent' (**Figure S4**). Therefore, the user-perceived agreement between PS and QR is likely higher than the concordance score indicates.

**Use Case 3: Renal tubular classification**

Tubules are a major component of the nephron, the functional unit of the kidney. The two major types of tubules in the kidney cortex are the proximal and distal tubules, and they are vulnerable to a variety of injuries across diseases (e.g., atrophy, acute injury, osmotic changes, etc.). For this use case, tubules were labeled into four classes: proximal, distal, abnormal, and other (*i.e.*, false positive from the *a priori* tubule segmentation step and collecting ducts or thin limb of loop of Henly tubules in the medulla).[25]

216 ROIs were extracted from fully deidentified WSI from the NEPTUNE[26] PAS WSI cohort at 20x Magnification and uploaded into PS. ROIs were stain normalized using the Vahadane stain normalization[20] implementation from StainTools. 10,129 Tubules were extracted into 256 by 256 pixel patches with a single tubule centered based on tubule annotations created in QuPath[27].

**Use Case 4: Renal glomerular classification**

Glomeruli, the filtration organelles of the kidney nephrons, can undergo a variety of morphologic changes. For this use case, we selected diseases where glomeruli can undergo segmental to global scarring. Glomeruli were labeled into 5 categories: globally sclerotic (GS), segmentally sclerotic (SS), non-sclerotic glomeruli (non-SS/GS), non-glomeruli (*i.e.* false positive from *a priori* glomeruli segmentation step) and uncertain (*i.e.* distinction between SS and GS is challenging by visual inspection).[28,29] The high complexity of these organelles consisting of various cell types, a capillary tuft, a mesangial stalk, a urinary space, and a capsule, and the high heterogeneity in image presentation of GS and SS glomeruli, allows for the showcasing PS's ability to provide improved labeling efficiency of complex objects. The reported accuracy is measured between GS, SS, non-GS/SS and non-glomeruli labels. Cases labeled as uncertain were excluded as their ambiguous nature would not lead to meaningful conclusions regarding the concordance between PS and QR.

For the experiment, 16,158 glomeruli from 241 fully deidentified NEPTUNE[26] and CureGN[30] PAS WSIs were used. Glomeruli were previously manually segmented using QuPath[27] and preprocessed into 256 by 256 pixel ROI extracted at 40x magnification, each containing a singular glomerulus centered in the ROI. ROI were normalized using Vahadane stain normalization[20] using the StainTools library. ROI and corresponding segmentation masks were uploaded into PS according to the PS workflow (**Figure S2**). Patches were created using the full ROI.

**Configuration and hyperparameters**

The default version of PS is nearly fully configured. The few hyperparameters of interest are easily modifiable through the configuration file. In the use cases discussed here, the hyperparameters requiring change relate to the patch size extracting the objects from the ROI images as well as the encoder size of the DL model, governing how much information for a given patch can be used by the model to assess patch similarities. Patch size was chosen based on object size and magnification, such that each object is fully visible in a patch. In the use cases presented (**Table 1**), the encoder size was set equal to the patch size. For example, in the glomeruli classification use case, patch size was configured as 256x256 pixels, with the encoder size being configured as 256. This parameter setting approach appears to yield a sufficient starting point for using PS efficiently.

**Experiment setup**

Each experiment was conducted on an Ubuntu Server 20.04LTS equipped with a Nvidia GeForce RTX 2080 Ti.

## Supplementary Figures

### Supplementary figure 1: PS enables direct workflow from object labeling to downstream analysis

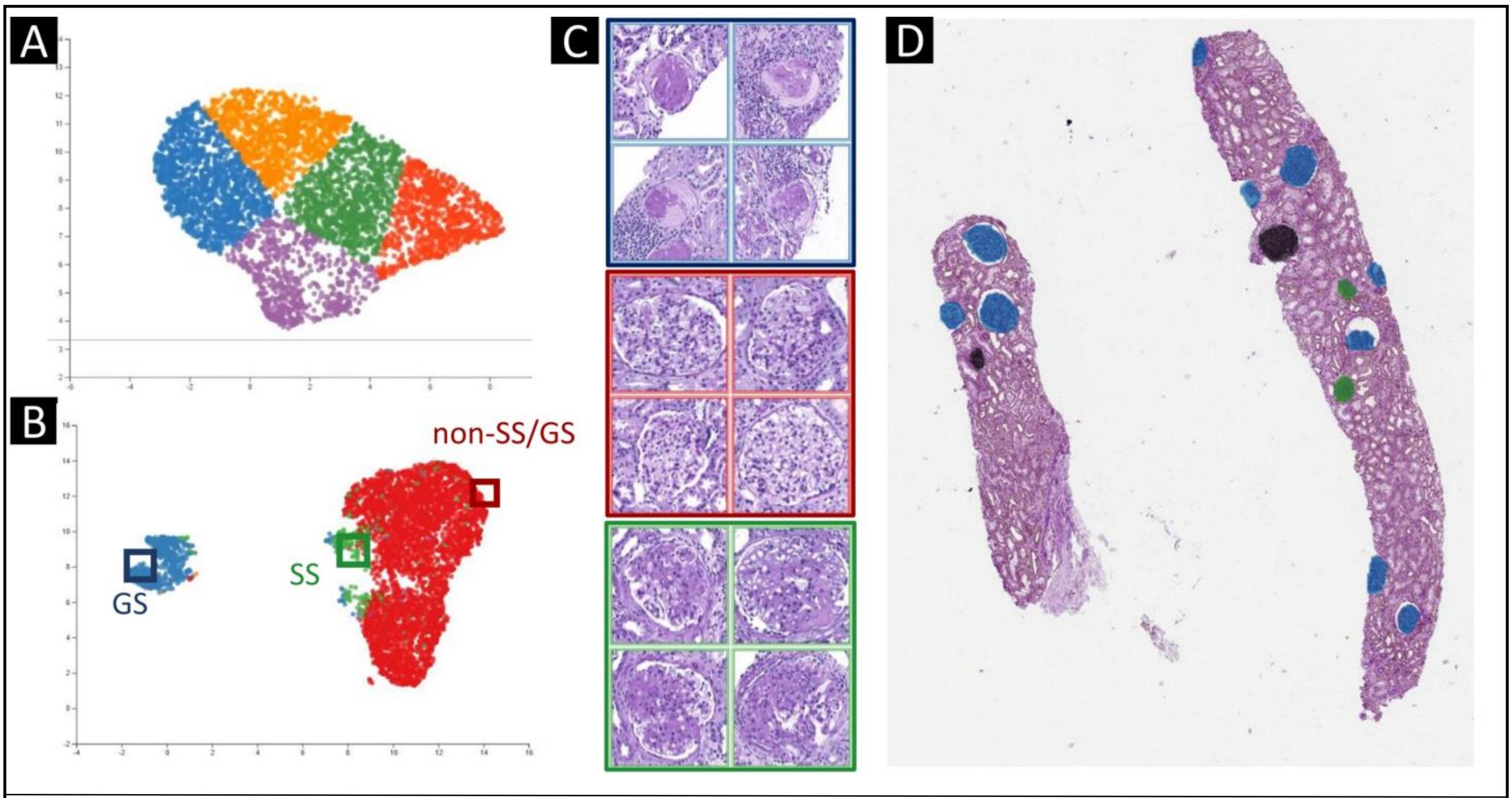

**Figure S1.** Initial unsupervised embedding space of the glomeruli classification use case (A) clusters the embedding space equally amongst the 5 classes. (B) After four training, embedding and labeling cycles, the embedding space shows clear class separation (blue, green, and red box), (C) yielding high class purity regions suitable for bulk labeling. As PS supports the extraction of objects from ROIs, if suitable ROIs are supplied, PS can generate ground truth and prediction masks directly usable for downstream analysis (D).

**Supplementary figure 2: PS workflow diagram**

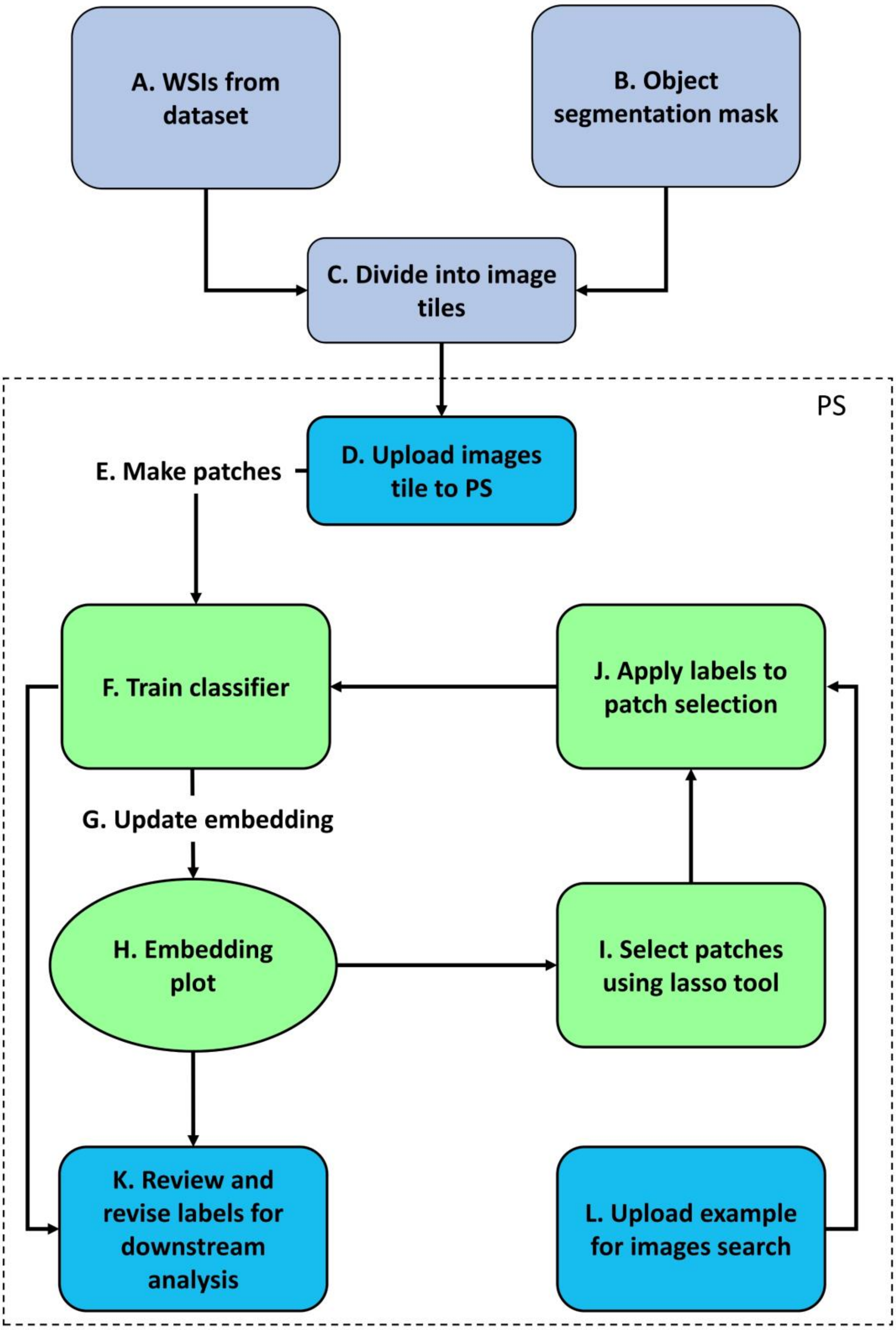

**Figure S2.** Flowchart illustrating the workflow of PS. (A & B & C) WSI are divided into tiles for uploading into PS according to the use case specific workflow (see **Online Methods**). (D) Images tiles are uploaded into PS where (E) tiles are then subsequently divided into smaller patches of use case specific size (Table S1) with the objects centered suitable for the deep learning model. (F) Patches are then used to train a deep learning model to encode patch information into a feature vector. (G) Each patch is subsequently embedded into a two-dimensional space using UMAP from the feature representation layer of the deep learning model. (H->I) From this the iterative cycle begins where the user selects patches from the UMAP embedding plot (Figure 1A) and assigns class labels to the selection (Figure 1A). Label information is incorporated into the DL training which increased class separation in the embedding space. (K) At any point, model predictions can be reviewed on an ROI level (Figure 1BC) and object labels can be revised directly on the output generation (Figure 1C), which can be used for further downstream analysis. (L) Instead of the lasso tool (H), the user can alternatively select patches from the embedding space using a reference image.

**Supplementary figure 3: Discordant cases in lymphocyte labeling in triple-negative breast cancer use case**

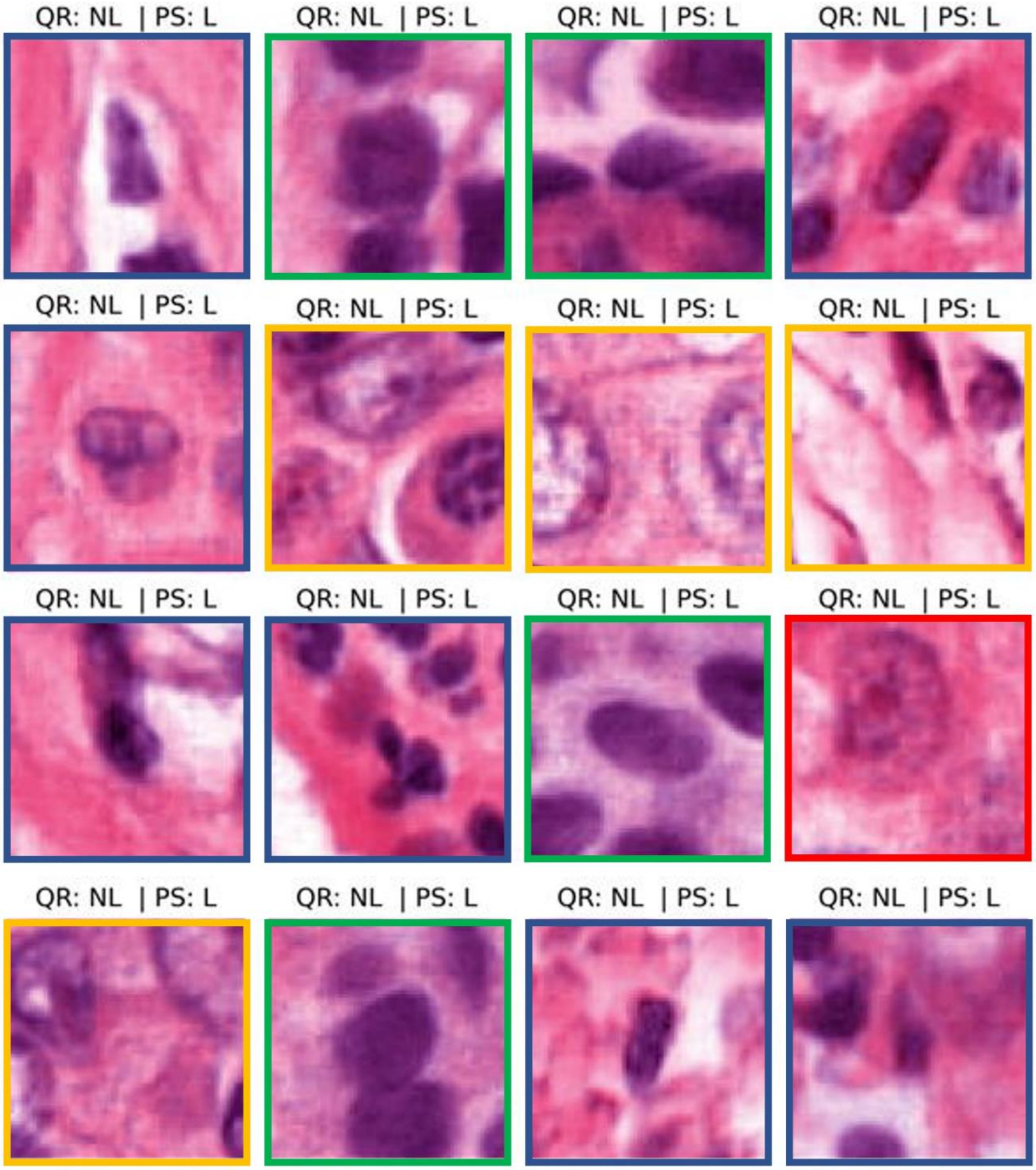

**Figure S3.** Discordant cases between PS and QR in the nuclei use case. All discordant objects were labeled as Lymphocyte (L) in PS and non-lymphocyte (NL) in QR. Discordant cases can be broadly distinguished into four categories. First, where no nucleus is present in the patch center due to nuclei segmentation errors (yellow box). As the user was forced to label the whole cohort, a decision for every patch had to be reached. In use cases which employ large scale automatic object detection, the inclusion of a general negative "non-object" class might be worth considering. Second, as the combined human labeling time exceeded 9 hours, human error is likely to occur for some patches (red box). Third, difficult exemplars with inter and intra observer variability (blue box), and fourth, label is context dependent (green box). As QR shows the whole ROI at once, context is immediately available, while in PS context information can be invoked manually per patch as deemed necessary by the user.

**Supplementary figure 4: Discordant cases in detection of tumor budding in pulmonary squamous cell carcinoma use case**

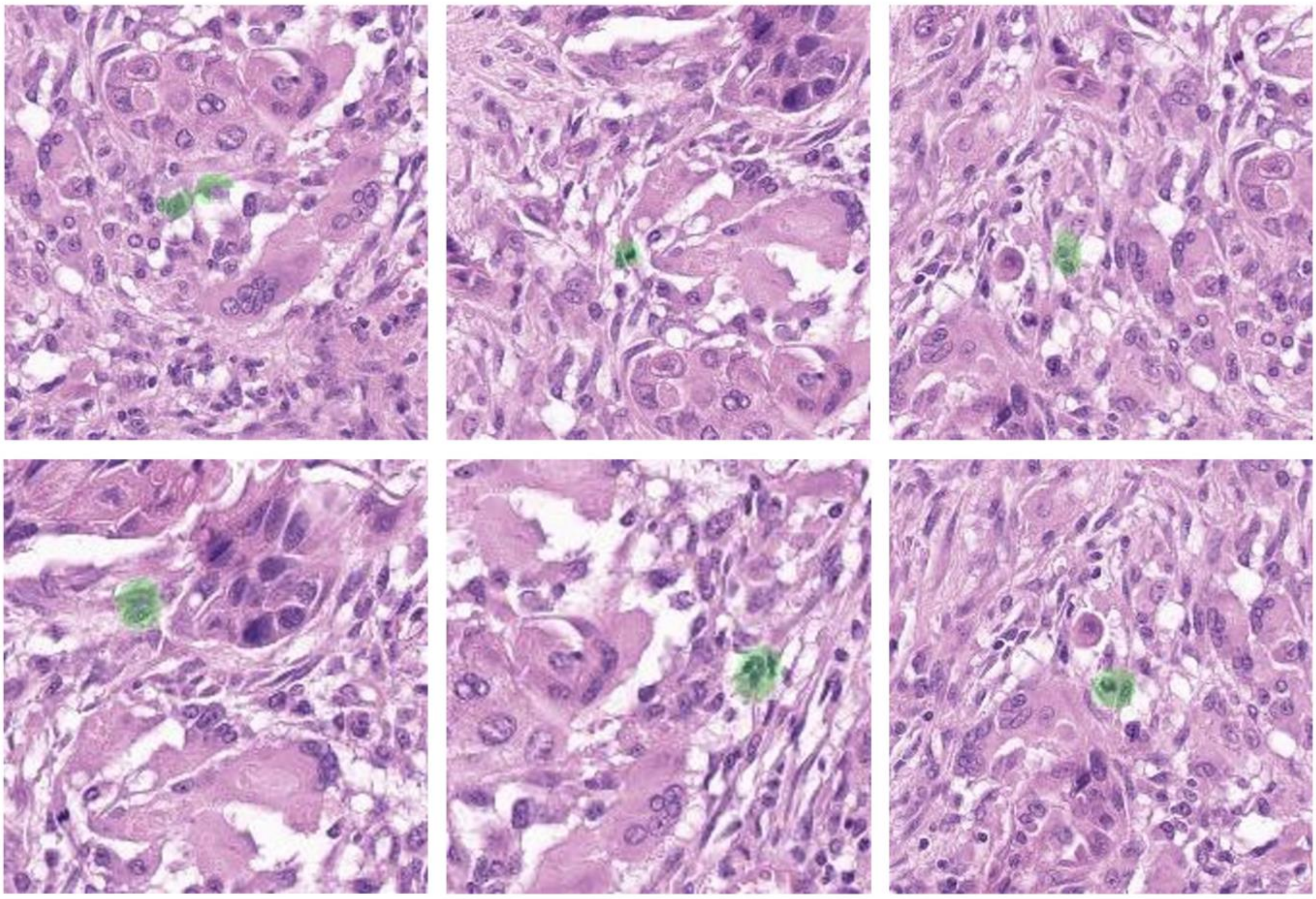

**Figure S4.** Discordant cases between PS and QR in the lung tumor budding use case with u-net generated segmentation mask indicating tumor-bud candidate position (green). This overlay was necessary as multiple tumor bud candidates may be present per ROI. All discordant cases shown were labeled as unsure in QR and absent in PS. QR, as opposed to PS, presents a ROI rather than a patch, making context immediately available to the user. Discussion of discordant patches with the pathologist conducting the experiment indicates that the availability of context led the pathologist to be more doubtful in labeling patches as 'absent'. In PS however, proximity of the patch in the embedding space to other 'absent' patches led the pathologist to label these patches as 'absent'.

**Supplementary figure 5: Discordant cases in renal tubular classification use case**

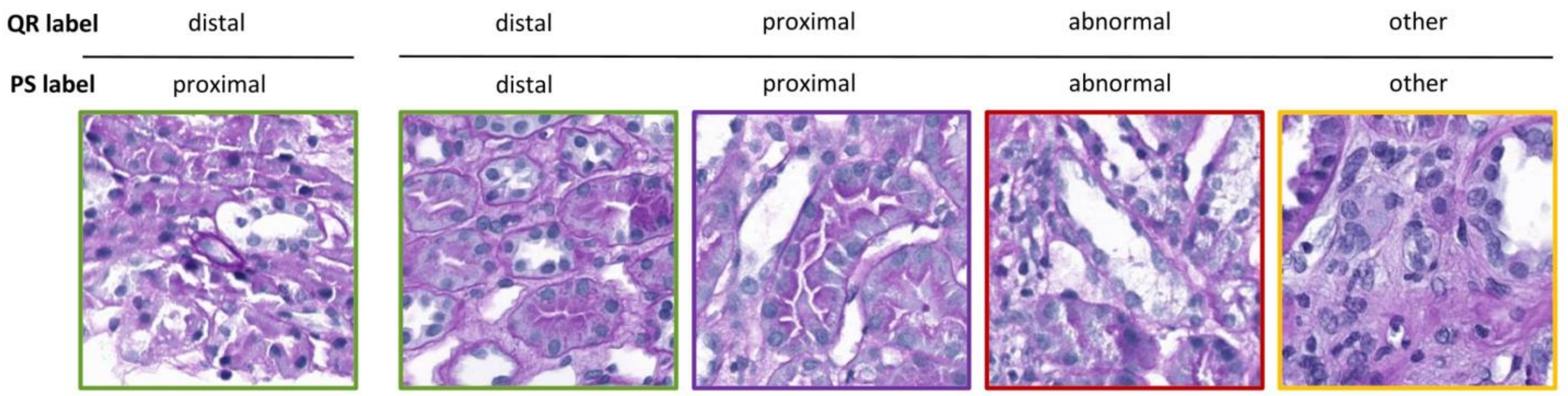

**Figure S5.** Discordant case between PS and QR in the renal tubule labeling use case (left) and four examples of tubules per labeled class (right), with color indicating QR labels: distal (green), proximal (purple), abnormal (red) and other (e.g., non-tubule) (yellow). Discordant case between PS and QR was labeled as proximal in PS, while labeled distal in QR. The discordant image contains a small proximal tubule in the center and a distal tubule just off-center. It is therefore likely that during labeling efforts this led to confusion on which object to label.

**Supplementary figure 6: Discordant cases in renal glomerular classification use case**

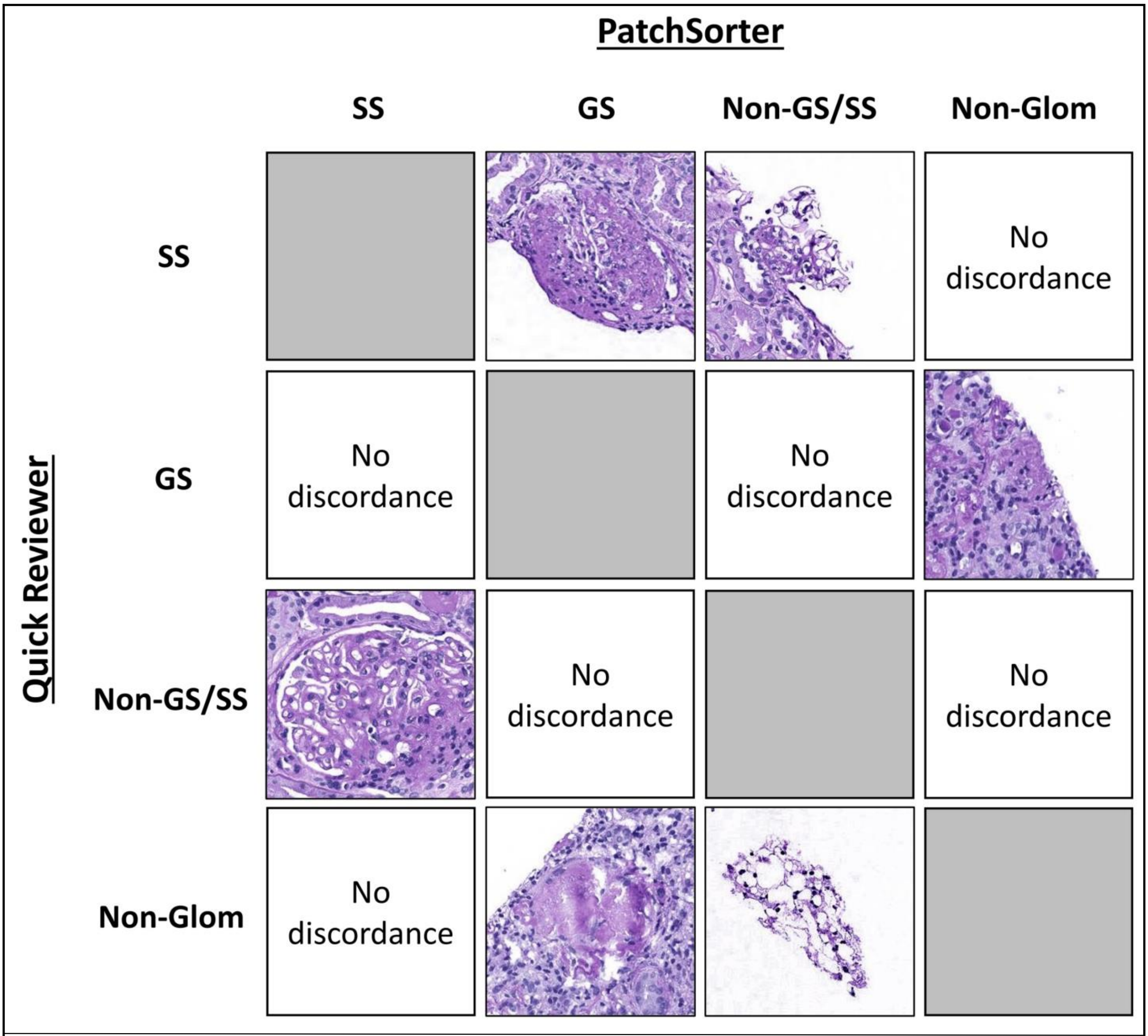

**Figure S6.** Discordant cases between PS and QR in the glomerular labeling use case. The structural complexity of the glomeruli and the high heterogeneity in SS and GS morphology, are likely contributing to the discordance between PS and QR. For example, a near to globally scarred glomerulus was labeled as SS using PS and vice versa (third panel from right). Similarly, a dense area of tissue is labeled GS on QR and non-glom on PS, and a glomerulus with a small area of sclerosis is labeled SS on QR and non-GS/SS on PS. Other examples of discordance labeling include 1) a glomerulus labeled non-GS/SS on QR and SS on PS: in this case the solidified area was interpreted as the vascular pole in one case and as sclerosis in the other; 2) a hyaline cast was labeled non-glom by QR but mistakenly labeled as GS on PS; 3) a floating fragment of non-better identifiable tissue (non-Glom by QR) is mistaken for non-GS/SS on PS.